\documentclass[aps,amssymb,amsmath,prl,preprint,noshowpacs]{revtex4-1}\fussy
\usepackage{amssymb,amsmath,graphicx}
\usepackage[usenames]{color}
\usepackage{bbold}
\usepackage{comment}

\newcommand{\commentOut}[1]{}

\usepackage{textcomp}
\newcommand{\murm}{\hbox{\textmu}}
\newcommand{\imu}{\text{\rm i}}
\newcommand{\figwidth}{\columnwidth} 

\usepackage{scalefnt}   

\begin{document}
\title{Nanomechanical method to gauge emission quantum yield applied to NV-centers in nanodiamond}
\author{Martin Frimmer}
\affiliation{Center for Nanophotonics, FOM Institute AMOLF,
Science Park 104, 1098 XG Amsterdam, The Netherlands}
\author{Abbas Mohtashami}
\affiliation{Center for Nanophotonics, FOM Institute AMOLF,
Science Park 104, 1098 XG Amsterdam, The Netherlands}
\author{A. Femius Koenderink}\email{koenderink@amolf.nl}
\affiliation{Center for Nanophotonics, FOM Institute AMOLF,
Science Park 104, 1098 XG Amsterdam, The Netherlands}




\begin{abstract}
We present a technique to nanomechanically vary the distance between a fluorescent source and a mirror, thereby varying the local density of optical states at the source position. Our method can therefore serve to measure the quantum efficiency of fluorophores. Application of our technique to NV defects in diamond nanocrystals shows that their quantum yield can significantly differ from unity. Relying on a lateral scanning mechanism with shear-force probe-sample distance control our technique is straightforwardly implemented in most state-of-the-art near-field microscopes.
\end{abstract}
\date{\today. Prepared for Applied Physics Letters}

\maketitle

Across fields ranging from solid state lighting, single molecule spectroscopy, optoelectronics, to quantum optics, nanophotonics is used to engineer light sources. Coupling single emitters to gratings, cavities, or subwavelength plasmonic antennas~\cite{Novotny2011} provides control over polarization~\cite{Moerland2008} and directionality~\cite{Curto2010} of single-photon emission at boosted emission rate~\cite{Kuehn2006,Novotny2006}. For all these applications emitters need to be well calibrated before device assembly to ensure that they meet  requirements regarding photostability, lifetime, absorption and emission spectrum. A property of paramount importance is the quantum yield  $QE=\gamma_\text{r}/\gamma_\text{tot}$, given as the ratio of radiative decay rate $\gamma_\text{r}$ and total decay rate $\gamma_\text{tot}=\gamma_\text{r}+\gamma_\text{nr}$, with $\gamma_\text{nr}$ the nonradiative rate~\cite{Novotny2006}. Importantly, $\gamma_r$ is proportional to the local density of optical states (LDOS), a property of the electromagnetic environment of the source~\cite{Novotny2006}. Therefore, exposing the emitter to a known LDOS variation while measuring its decay rate is an unambiguous way to determine the source's quantum efficiency~\cite{Buchler2005,Leistikow2009,Chizhik2011}. Buchler \emph{et al.} have pioneered a micromechanical method to vary the LDOS by approaching a mirror to a single molecule. Similarly, Chizhik \emph{et al.} made use of a nanomechanically tunable cavity~\cite{Chizhik2011}.

This Letter reports on a variation of the  nanomechanical technique of Buchler et al.~\cite{Buchler2005} to calibrate decay constants of fluorophores using the LDOS.  As in Ref.~\cite{Buchler2005} we use a spherical silver mirror of several 10\,\murm m in diameter attached to a scanning probe. Instead of retracting the mirror,  our technique relies on the shear-force mechanism to keep the mirror in near-contact with the sample and a lateral scanning procedure is used that is  implementable with most state-of-the-art near-field microscopes. We calibrate our technique both on colloidal beads infiltrated with high-quantum-yield dye and on thin emissive layers, as frequently encountered in organic light emissive devices. Finally we use our technique to measure the quantum efficiency of single NV defects in diamond nanocrystals. Such NV centers are considered highly promising  for indefinitely stable solid-state single-photon sources~\cite{Kurtsiefer2000, Kolesov2009,Schietinger2009,Huck2011}. A  problematic issue is that such defect centers, even though intrinsically identical quantum systems, are known to have different brightnesses and rates  due to differences  in their  embedding host.
Our measurements show that the quantum yield of NV defects in diamond nanocrystals can differ significantly from unity and varies between nanocrystals.

Our nanomechanical technique of controlling the LDOS  relies on moving a micron-sized mirror attached to a scanning probe, as in Ref.~\cite{Buchler2005}. Our scheme of changing the distance between a fluorophore and a mirror is illustrated in Fig.~\ref{Fig:setup}(a). The interrogated source is fixed to a substrate and a large spherical mirror is laterally scanned across the sample surface while the mirror-sample distance is kept constant using shear-force feedback~\cite{Novotny2006}.  This procedure is implementable in most scanning probe microscope, as opposed to calibrated retraction of the probe. In Fig.~\ref{Fig:setup}(a) two positions of the mirror with respect to an emitter are shown to illustrate the principle of changing the emitter-mirror distance by laterally moving the mirror over the sample.
To fabricate the micromirror we glue polystyrene beads (diameter 25\,\murm m, Polysciences Europe) to the cleaved end of an optical fiber with a small amount of super-glue. We subsequently evaporate a layer of 400\,nm of Ag onto the sphere to obtain a spherical micromirror. The optical fiber is then super-glued to a quartz tuning fork, as sketched in Fig.~\ref{Fig:setup}(b, inset), which is used as a shear-force probe.
{\begin{figure}
\includegraphics[width=\figwidth]{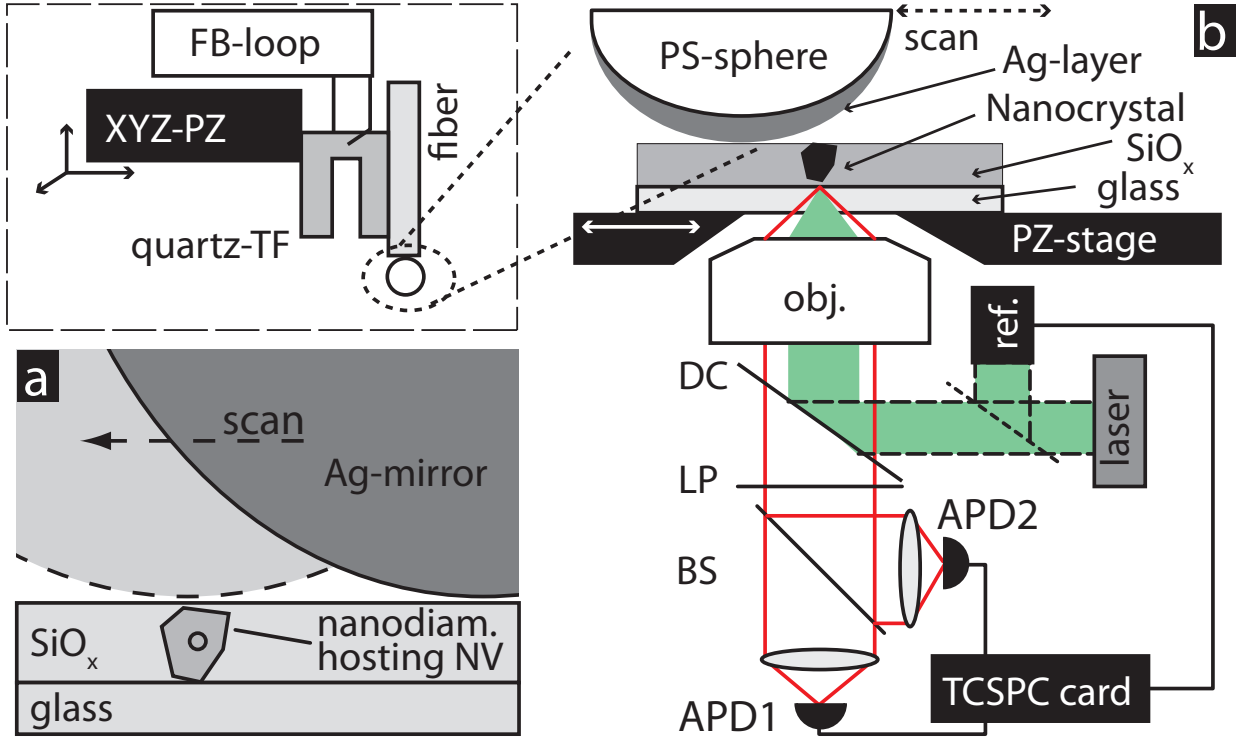}
\caption{(a)~Experimental principle: The distance between an  emitter  and a mirror is varied by laterally scanning a spherical mirror over the sample.
(b)~Sketch of experimental setup. The sample lies on a piezo  stage with the micromirror located above. The mirror is attached to an optical fiber glued to a quartz tuning fork, which is positioned with an xyz-piezo scanner [see inset]. Below the sample a microscope objective focuses the pump laser on the emitter. Fluorescence  is filtered by a dichroic beamsplitter (DC) and  long-pass filter (LP) before it is guided to a Hanbury Brown-Twiss setup composed of a beamsplitter (BS) and two APDs.}
\label{Fig:setup}
\end{figure}
}

To calibrate our nanomechanical technique of changing the LDOS we prepare a sample of dye-doped polystyrene beads (diameter 100\,nm, F8801, Invitrogen) dispersed at low concentration on a cleaned glass coverslip, such that individual beads are separated by several microns. We evaporate about 60\,nm of SiO$_2$ on top of the sample for mechanical protection. On our inverted confocal fluorescence-lifetime imaging (FLIM) microscope [sketched in Fig.~\ref{Fig:setup}(b), see Ref.~\onlinecite{Frimmer2011} for details] we locate a single fluorescing bead in the focus of a picosecond pump laser (532\,nm, 10\,MHz) generated by the objective (100$\times$, NA 1.4), which is confocally imaged onto an avalanche photodiode (APD). The APD is connected to a timing card to record the fluorescence lifetime of the source using standard time-correlated single-photon counting (TCSPC). With the fluorescing bead continuously in focus, we now use the scanning head of the setup to approach the micromirror to the sample surface, bring it into shear-force contact~\cite{Novotny2006} and raster-scan it laterally across the sample. In Fig.~\ref{Fig:technique}(a) we show a fluorescence-intensity map obtained while scanning the mirror, where each pixel corresponds to a certain horizontal position of the mirror with respect to the dye-doped bead, which leads to a certain emitter-mirror distance due to the curvature of the mirror, as seen in Fig.~\ref{Fig:setup}(a). We interpret the pronounced brightness variations observed in Fig.~\ref{Fig:technique}(a) as a result of the varying pump field experienced by the dye-doped bead due to interference of the incoming field and its reflections at both the mirror surface and the substrate-air interface.
{
\begin{figure}
\includegraphics[width=\figwidth]{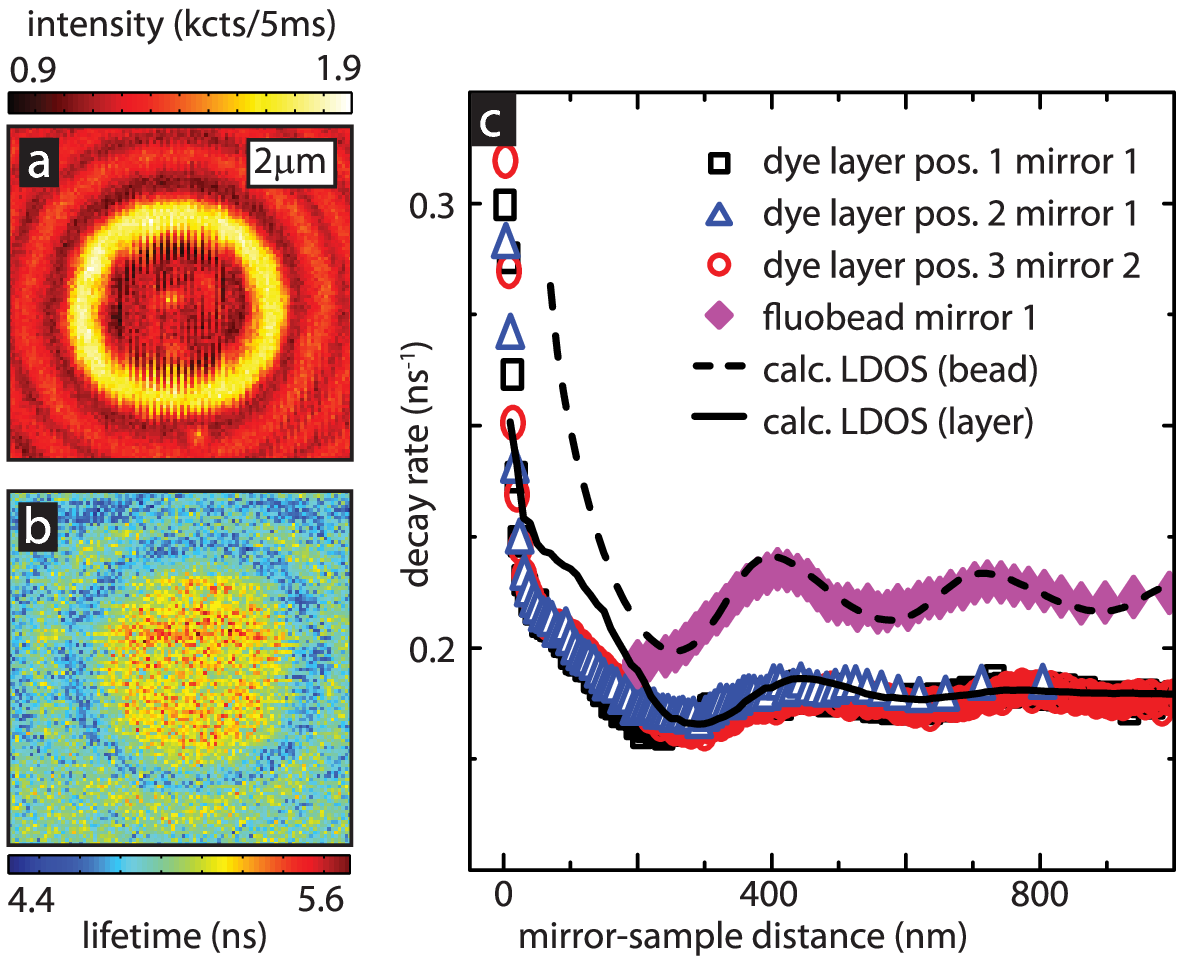}
\caption{(a)~Fluorescence-intensity map of dye-doped bead in laser focus as micromirror is scanned across it, showing ring-shaped intensity variations. Each pixel in the map denotes a specific lateral position of the mirror.
(b)~Fluorescence-lifetime map of same measurement that yielded~(a).
(c)~Decay rate as function of mirror-sample distance, as obtained from lifetime map shown in~(b) by combining pixels with identical distance to the center of the circular pattern. Full diamonds are data obtained from (b), measured on 100\,nm fluorescing bead. Open symbols show results on continuous dye layers (different positions, and different micromirrors [see legend]). Solid and dashed lines are analytical calculations for dye layer and fluorescing bead, respectively.}
\label{Fig:technique}
\end{figure}
}
Analysis of the arrival times of the fluorescent photons collected at different mirror positions yields the FLIM map shown in Fig.~\ref{Fig:technique}(b), exhibiting circularly symmetric variations in the detected lifetime. The circular symmetry of both Figs.~\ref{Fig:technique}(a) and (b) is the expected result of the symmetry of our micromirror with respect to its touching point with the sample. We exploit this symmetry by determining the center of the patterns in Figs.~\ref{Fig:technique}(a,b) and combining TCSPC data of pixels with equal distance to the center of the pattern into a single decay trace. When we furthermore use the known geometry of the micromirror to convert in-plane distances  in Fig.~\ref{Fig:technique}(b) into a mirror-emitter separation, we obtain the decay rate of the emitters as a function of their distance to the mirror surface, shown as the full diamonds in Fig.~\ref{Fig:technique}(c). We clearly observe characteristic decay-rate oscillations in front of the mirror~\cite{Drexhage1970}.

To  model our experiments, we consider an air layer sandwiched between two semi-infinite half-spaces, the upper one being Ag ($\epsilon=-15.5+0.52\imu$ at 620\,nm, measured by ellipsometry on an Ag film on a Si substrate), the lower one being glass ($\epsilon=2.25$) . We thereby approximate the spherical mirror as a plane, an assumption that holds whenever the distance between the emitter and the sphere is much smaller than the sphere's radius of curvature. We calculate the LDOS in a range of depths into the glass substrate corresponding to the fluorescing bead diameter using the methods in Refs.~\onlinecite{Paulus2000,Novotny2006}. We assume an emission wavelength of 620\,nm, the emission maximum of the fluorescing beads. Subsequently, we average the obtained enhancement factors for different emitter orientations assuming a homogeneously distributed ensemble of emitters, since a bead contains about $10^3$ dye molecules~\cite{Frimmer2011}. The case of the dye-doped bead covered with an evaporated SiO$_x$ layer is hard to model, since the protruding bead breaks the analytically accessible planar geometry~\cite{Paulus2000,Novotny2006}. We approximate the situation with a stratified medium, shifting the mirror-sample distance axis by 60\,nm in Fig.~\ref{Fig:technique}(c) to accommodate the height of the SiO$_x$ layer. The obtained correspondence between measured data [full diamonds in Fig.~\ref{Fig:technique}(c)] and calculation (dashed curve, assuming unit quantum efficiency) is fully satisfying.

While using a dye-doped bead for a reference measurement has the virtue of the source being confined to a subwavelength volume, the protective capping layer limits how close the mirror can get to the emitters. As a second test, we employ an alternative geometry, where we spin coated a dye-doped PMMA layer of about 70\,nm thickness onto a glass coverslip and repeated the experiment shown in Figs.~\ref{Fig:technique}(a,b).
To this end we diluted 5\,mg dye (Bodipy TR, D-6116, Invitrogen) in 1\,ml anisole to further dilute that mixture 30-fold in a 2\% mixture of PMMA with molecular weight 950K in anisole. The PMMA mixture was then spun on cleaned coverslips and baked for 5\,minutes at 180$^\circ$C. We note that the manufacturer does not disclose the precise type of dye  in the fluorescing beads used for the measurements in Fig.~\ref{Fig:technique}(c). Comparison of spectrum and lifetime strongly suggests that the dye incorporated in the beads is Bodipy TR, which we use throughout this paper to prepare dye-doped PMMA layers.
Using a continuous dye-doped layer the spatial selection in the sample plane is solely  through the pump spot, which is about 500\,nm diameter in our case. The resulting measured decay rate as a function of mirror-sample distance is shown as the black squares in Fig.~\ref{Fig:technique}(c).
A distance of zero on the horizontal axis corresponds to the mirror being exactly above the interrogated point on the sample in the experiment and the size of the air-gap being zero in the calculation. We have repeated the experiment of probing the lifetime in a dye layer [open squares in Fig.~\ref{Fig:technique}(c)] both with the same micromirror at a different location on the same sample [open triangles] and with a different micromirror [open circles]. The resulting lifetime traces agree excellently, showing that our technique of modifying LDOS by nanomechanical control of a micromirror is both fully reversible and repeatable, even with different micromirrors.
When comparing the data for the dye-doped bead [full diamonds in Fig.~\ref{Fig:technique}(c)] with the data obtained on the continuous  layer [open symbols], one immediately notes the larger decay rate of the emitters in the  bead. Assuming that the molecular dye is indeed identical, the rate of the bead is expected to be enhanced since it is entirely surrounded by high-index SiO$_x$ while the dye layer is in immediate proximity to the air half-space. Furthermore, we note the reduced contrast in the decay-rate variations of the continuous layer as compared to the bead. This reduced contrast is a result of the larger lateral extent of the probed volume. The focal size of the pump spot (diameter  500\,nm) results in a broader distribution of distances of probed emitters from the mirror, an effect becoming increasingly severe at larger emitter-mirror separations due to the mirror's curvature. We model the experiment on the dye layer by averaging the decay-rate enhancements experienced by emitters located in a volume given by the dye-layer thickness and the focal spot-size. While the obtained analytical result, plotted as the solid line in Fig.~\ref{Fig:technique}(c), is fully satisfactory at larger sample-mirror distances, the correspondence between theory and experiment is only qualitative for mirror-sample separations smaller than about 200\,nm. There are several possible reasons for the observed discrepancy. First, effects like mirror-surface roughness  could have an effect at these small distances. Furthermore, the fact that we are probing an ensemble complicates things, since varying detection efficiencies for differently oriented or positioned emitters  makes a selection of a subensemble of emitters that varies with mirror-sample distance~\cite{Lukosz1979}. As a check, we have performed measurements on a dye layer with annular illumination, where the pump-field is polarized preferentially perpendicular to the sample surface~\cite{Buchler2005}. Since we found no significant effect of the illumination conditions on the lifetime we exclude a selection of a subensemble by the pump field. Accordingly, we attribute the discrepancy between theory and experiment at small mirror-sample separations to variations in the collection efficiency for different emitter subensembles.

\begin{figure}
\includegraphics[width=\figwidth]{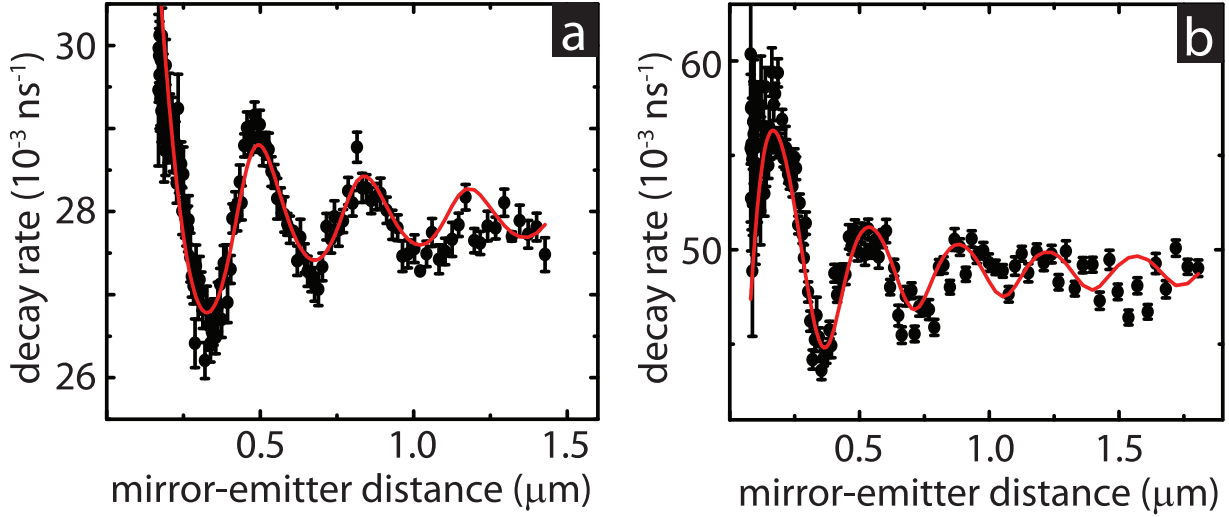}
\caption{(a) Data-points: Decay rate of a single NV center in a diamond nanocrystal as a function of distance to the scanning mirror. Error bars denote $1\sigma$-interval. Red solid line: Calculation of LDOS experienced by emitter at depth $d=150\,\text{nm}$ into the substrate, with dipole moment in the sample plane ($\theta=0^\circ$) and $QE=26\%$. (b) Same as (a) but for NV defect in different nanocrystal (Calculation:   $QE=58\%$). }
\label{Fig:NV_drex}
\end{figure}
We now apply our technique of nanomechanically changing the LDOS to NV defect centers in diamond nanocrystals. We prepare samples of diamond nanocrystals (Microdiamant, MSY0--0.2, median diameter 108\,nm, $<1\%$ of particles larger than 175\,nm) by diluting the stock solution 1:100 before spin coating it on cleaned quartz coverslips. We cover the sample with a 200\,nm layer of spin-on glass (FOX-14, Dow Corning, n=1.4) for mechanical protection. All measurements on diamond nanocrystals are taken with scanning mirrors with a diameter of 45\,\murm m and a dry objective (NA 0.9).  A detailed statistical study on many nanocrystals that we report elsewhere identifies fewer than 0.1\% of the nanocrystals to host a single NV center, identified  through the characteristic zero-phonon line around 637\,nm  and $g^2(\tau=0)<0.5$~\cite{abbaslong}.   Having identified a nanocrystal hosting an NV-center we now apply our scanning-mirror method to obtain a measurement of its fluorescence lifetime as a function of its distance to the mirror, as shown in Fig.~\ref{Fig:NV_drex}(a). We observe a clear variation of the decay rate of the NV-center between about 0.026\,ns$^{-1}$ to 0.030\,ns$^{-1}$, which we attribute to the varying LDOS in front of the mirror. A second dataset obtained from a different nanocrystal hosting an NV center is shown in Fig.~\ref{Fig:NV_drex}(b).

We now turn towards interpreting the measurement in Fig.~\ref{Fig:NV_drex}. To be inferred from a fit to the data are the emitter's dipole moment orientation, described by the angle $\theta$ between dipole moment and surface ($0\le\theta\le90^\circ$), its quantum efficiency $QE$ ($0\le{QE}\le1$), and  its total decay rate $\gamma_\text{tot}$ in absence of the mirror.  We sweep the parameter  $\theta$ through its range  and at each value of  $\theta$ apply a maximum likelihood fitting routine to obtain the values of $QE$ and $\gamma_\text{tot}$, as well as the residuals between the fitted curve and the measured dataset. By minimizing the residuals of the fit we obtain the set of most likely parameters. For the NV center that yielded the dataset shown in Fig.~\ref{Fig:NV_drex}(a) we find as most likely values $\theta=0^\circ$ and ${QE}=26\%$ for which we plot the analytical result as the red solid line in Fig.~\ref{Fig:NV_drex}(a). Likewise, for the NV center studied in Fig.~\ref{Fig:NV_drex}(b) we obtain $\theta=0^\circ$ and ${QE}=58\%$.
Furthermore, by comparing the residuals of the fit to the uncertainty of the measurement we are able to identify the confidence interval beyond which the parameters are incommensurate with the data given the measurement uncertainty. We adopted this cautious procedure due to the strong cross correlation between QE and $\theta$. The range of $\theta$ commensurate with the measurement uncertainty of the data in Fig.~\ref{Fig:NV_drex}(a) is $\Delta\theta=\pm45^\circ$. Even within that range of $\theta\pm\Delta\theta$, the quantum efficiency remains bounded to ${QE}\le50\%$. We therefore conclude that the quantum efficiency of the individual NV defect of Fig.~\ref{Fig:NV_drex}(a) does not exceed 50\% and is most likely as low as 26\%. Along the same lines we can conclude that the quantum yield of the NV center in Fig.~\ref{Fig:NV_drex}(b) cannot exceed 90\%.

From our data we conclude that the quantum yield of NV centers varies strongly from nanocrystal to nanocrystal. This finding might seem surprising given that near unity quantum efficiency has been well established for NV centers in bulk~\cite{Gruber1997} and is commonly assumed for NV defects in nanocrystals as well~\cite{Schietinger2009,Huck2011}. However, we note that NV centers in nanocrystals are  inherently close to the nanocrystal surface, which might provide nonradiative recombination pathways. Our findings call into question the suitability of diamond nanocrystals as localized probes of their electromagnetic environment. While it has been well known that  NV centers in nanocrystals exhibit a large spread in lifetimes~\cite{Kolesov2009,Ruijgrok2010}, we have established that there also exists a spread in quantum efficiencies. Our findings imply that only an individually characterized NV center in a diamond nanocrystal can serve as a calibrated local LDOS probe.

In conclusion, we have presented a technique to nanomechanically vary the LDOS experienced by a quantum emitter in a controlled fashion. Our technique relies on laterally scanning a micromirror attached to a scanning probe held at a constant height above the substrate using the shear-force mechanism. Our approach is easily implemented in practically all near field microscopes. We used our technique to gauge the quantum efficiency of individual NV defect centers in diamond nanocrystals. We established that NV centers in nanocrystals can exhibit a quantum yield as low as 25\% in stark contrast to common belief. Accordingly, we advocate careful precharacterization of NV defects in nanocrystals before their application as LDOS probes. Our technique offers an easily implementable solution for such characterization and is readily extendable to other emitters that suffer from ensemble variations, such as colloidal semiconductor quantum dots~\cite{Leistikow2009}.

\begin{acknowledgments}
This work is part of the research program of the ``Stichting voor Fundamenteel
Onderzoek der Materie (FOM)'', which is financially supported by the
``Nederlandse Organisatie voor Wetenschappelijk Onderzoek (NWO)''.  AFK gratefully acknowledges a NWO-Vidi grant for financial support.
\end{acknowledgments}

\end{document}